\documentclass[journal=cmatex,manuscript=article]{achemso}

\usepackage[version=3]{mhchem}
\usepackage{graphicx}
\usepackage{amsmath}
\usepackage{multirow}
\usepackage{color}
\usepackage{soul}
\usepackage[hidelinks]{hyperref}
\setkeys{acs}{articletitle = true}
\SectionNumbersOn
\AbstractOn
\title{Role of point defects in spinel 
Mg chalcogenide conductors
}

\author{Pieremanuele Canepa} \email{pcanepa@lbl.gov}
\affiliation{
Materials Science Division, 
Lawrence Berkeley National Laboratory, Berkeley, CA 94720, USA}
\altaffiliation{Current address: Department of Chemistry, University of Bath, Bath, BA2 7AY, UK.}
\alsoaffiliation{Contributed equally to this work}

\author{Gopalakrishnan Sai Gautam}
\affiliation{
Department of Materials Science and Engineering, Massachusetts
Institute of Technology, Cambridge, MA 02139, USA}
\alsoaffiliation{
Materials Science Division, 
Lawrence Berkeley National Laboratory, Berkeley, CA 94720, USA}
\alsoaffiliation{Contributed equally to this work}

\author{Danny Broberg}
\affiliation{Department of Materials Science and Engineering, University 
of California Berkeley, Berkeley, CA 94720, USA}

\author{Shou-Hang Bo}
\affiliation{
Materials Science Division, 
Lawrence Berkeley National Laboratory, Berkeley, CA 94720, USA}

\author{Gerbrand Ceder}  \email{gceder@berkeley.edu}
\affiliation{
Materials Science Division, 
Lawrence Berkeley National Laboratory, Berkeley, CA 94720, USA}
\alsoaffiliation{
Department of Materials Science and Engineering, Massachusetts
Institute of Technology, Cambridge, MA 02139, USA}
\alsoaffiliation{Department of Materials Science and Engineering, University 
of California Berkeley, Berkeley, CA 94720, USA}

\begin{document}
%\linenumbers
%%%%%%%%%%%%%%%%%%%%%%%%%%%%%%%%%%%%%%%%%%%%%%%%%%%%%%%%%%%%%%%%%%%%%%%%

%%%%%%%%%%%%%%%%%%%%%%%%%%%%%%%%%%%%%%%%%%%%%%%%%%%%%%%%%%%%%%%%%%%%%%%%
% abstract
%%%%%%%%%%%%%%%%%%%%%%%%%%%%%%%%%%%%%%%%%%%%%%%%%%%%%%%%%%%%%%%%%%%%%%%%
\begin{abstract}
Close-packed chalcogenide spinels, such as MgSc$_2$Se$_4$, MgIn$_2$S$_4$ and MgSc$_2$S$_4$, show potential as solid electrolytes in Mg batteries, but are affected by non-negligible electronic conductivity, which contributes to self-discharge when used in an electrochemical storage device. Using first-principles calculations,  we evaluate the energy of point defects as function of  synthesis conditions and Fermi level to identify the origins of the undesired electronic conductivity. Our results suggest that Mg-vacancies and Mg-metal anti-sites (where Mg is exchanged with Sc or In) are the dominant point defects that can occur in the systems under consideration. While we find anion-excess conditions and slow cooling to likely create conditions for low electronic conductivity, the spinels are likely to exhibit significant $n$-type conductivity under anion-poor environments, which are often present during high temperature synthesis. Finally, we explore extrinsic aliovalent doping to potentially mitigate the electronic conductivity in these chalcogenide spinels. The computational strategy is general and can be easily extended to other solid electrolytes (and electrodes) to aid in the optimization of the electronic properties of the corresponding frameworks.  
\end{abstract}
%%%%%%%%%%%%%%%%%%%%%%%%%%%%%%%%%%%%%%%%%%%%%%%%%%%%%%%%%%%%%%%%%%%%%%%%

%%%%%%%%%%%%%%%%%%%%%%%%%%%%%%%%%%%%%%%%%%%%%%%%%%%%%%%%%%%%%%%%%%%%%%%%
\section{Introduction}
\label{sec:intro}
%%%%%%%%%%%%%%%%%%%%%%%%%%%%%%%%%%%%%%%%%%%%%%%%%%%%%%%%%%%%%%%%%%%%%%%%
Chalcogenide materials, based on sulfur, selenium and tellurium, are used in a range of technological applications, including thermoelectric materials,\cite{Hsu2004,SnyderToberer2008}  semiconductors for light adsorbents and electronics,\cite{PerssonZhaoLanyEtAl2005,TodorovReuterMitzi2010,ChenWalshLuoEtAl2010,SunChanKangEtAl2011,SunCeder2011,RedingerBergDaleEtAl2011,BaranowskiZawadzkiChristensenEtAl2014,KitchaevCeder2016,Butler_2016} superconductors,\cite{PatrieFlahautDomage1964,GuittardSouleauFarsam1964,GambleDiSalvoKlemmEtAl1970,Mattheiss1973,WilliamsMcQueenCava2009}  Li-ion battery materials,\cite{Whittingham1978,Levi2009a,KamayaHommaYamakawaEtAl2011,OngMoRichardsEtAl2013,WangRichardsOngEtAl2015,CanepaBoGopalakrishnanEtAl2016}  quantum-dots,\cite{Nozik2002,EislerSundarBawendiEtAl2002,Talapin2005,AnikeevaHalpertBawendiEtAl2009,MeinardiColomboVelizhaninEtAl2014} and more recently, topological insulators.\cite{JiaJiCliment-PascualEtAl2011,ChhowallaShinEdaEtAl2013} Specifically, sulfides have already seen applications as solid electrolytes (or super-ionic conductors) in solid-state Li-ion batteries.{\cite{KamayaHommaYamakawaEtAl2011,OngMoRichardsEtAl2013,WangRichardsOngEtAl2015}}  The chalcogenide defect  chemistry, either in terms of intrinsic point defects or extrinsic substitutional impurities, has often been deemed responsible for their respective figures of merit.\cite{Behrh2017,Griffin2017}

Recently, ternary Mg-chalcogenide spinels were also identified as possible high mobility Mg-conductors.{\cite{CanepaBoGopalakrishnanEtAl2016}} This is relevant for the possible development of Mg transport coatings or solid state  electrolytes for Mg batteries,{\cite{CanepaBoGopalakrishnanEtAl2016}} which have the potential to outperform Li-ion batteries in terms of energy density.{\cite{Canepa2017}} The good Mg conductivity in the MgSc$_2$Se$_4$, MgIn$_2$S$_4$ and MgSc$_2$S$_4$ spinels is, however, plagued by non-negligible electronic conductivity.\cite{CanepaBoGopalakrishnanEtAl2016} Though the significant Mg ionic conductivity $\sigma_{ionic}\sim$~0.1~mS~cm$^{-1}$  (at 298 K) observed in MgSc$_2$Se$_4$ (\emph{via} $^{25}$Mg magic angle spin solid-state NMR and AC impedance spectroscopy) the electronic conductivity of MgSc$_2$Se$_4$ is $\sim$ 0.04 \% of the ionic conductivity,{\cite{CanepaBoGopalakrishnanEtAl2016}}  and substantially larger than in other state-of-the-art alkali (Li and Na) ion conductors ($\sigma_{electronic}$/$\sigma_{ionic}$ ~ 10$^{-4}$--10$^{-6}$ \%).{\cite{KamayaHommaYamakawaEtAl2011}}  Analogous to studies in semi-conductor applications,{\cite{FreysoldtGrabowskiHickelEtAl2014}} both intrinsic and extrinsic structural defects can cause large variations in electron (hole) conductivity in ionic conductors. Thus, we explore the defect chemistry of MgSc$_2$Se$_4$, MgSc$_2$S$_4$ and MgIn$_2$S$_4$ using first-principles calculations and aim to understand how structural defects modulate the electronic properties in the bulk spinels, identify the origin of the undesired electronic conductivity, and propose practical remedies.

In detail, our calculations suggest that intrinsic point defects, such as  Sc$^{3+}$ substituting on Mg$^{2+}$ sites in MgSc$_2$S$_4$ or MgSc$_2$Se$_4$ (Sc$^{\bullet} _{\rm Mg}$ using the Kr\"{o}ger--Vink notation), and In$^{\bullet}_{\rm Mg}$ and Mg$^{'} _{\rm In}$ in MgIn$_2$S$_4$,  can give rise to significant electronic conductivity in these materials.  Additionally, our data demonstrates that anion-rich and anion-poor synthesis conditions should give rise to qualitatively different defects, affecting the electronic (hole) conductivity of these materials. Finally, we demonstrate that understanding and controlling the defect chemistry of solid electrolytes (and cathode materials) is crucial in all aspects, such as tuning the respective synthesis conditions and optimizing the electronic and ionic conductivities. 
 %%%%%%%%%%%%%%%%%%%%%%%%%%%%%%%%%%%%%%%%%%%%%%%%%%%%%%%%%%%%%%%%%%%%%%%%
 \section{Methodology}
  %%%%%%%%%%%%%%%%%%%%%%%%%%%%%%%%%%%%%%%%%%%%%%%%%%%%%%%%%%%%%%%%%%%%%%%%
\subsection{Basics of defect chemistry}
The occurrence of a defect $X$ of charge  $q$ in a solid relates to its formation energy $E^f[X^q]$:
\begin{equation}{\small
E^f[X^q] = E_{tot}[X^q] - E_{tot}[bulk]  -\sum_i n_i \mu_i + qE_{Fermi}+ E_{corr}
}
\label{eq:defecformation}
\end{equation}

\noindent where,  $E_{tot}[X^q]$ and $E_{tot}[bulk]$ are the total energies of a supercell containing the defect $X$ and an un-defected supercell, respectively.{\cite{Zhang1991,Laks1992,Kwak1995,VandeWalle2004,FreysoldtGrabowskiHickelEtAl2014}} $n_i$ is the concentration of species $i$ added  ($n_i~>$~0)  or removed  ($n_i~<$~0)  to create defect $X$. $\mu_i$ is the chemical potential of species $i$, as determined by the set of phases in thermodynamic equilibrium with the solid of interest at 0~K. $E_{Fermi}$ is the Fermi energy of electrons in the structure, and $E_{corr}$ is the electrostatic correction term to account for spurious interactions among defects (i.e., with periodic images and the homogeneous background charge). Using Eq.~\ref{eq:defecformation}, the defect formation energies  $E^f[X^q]$ can be plotted as a function of the Fermi energy, $E_{\rm Fermi}$,  as demonstrated in Figure~\ref{fig:MgScSedefects}. 

In this work, we compute $E_{corr}$ using the Freysoldt correction scheme,\cite{FreysoldtNeugebauerWalle2009,FreysoldtNeugebauerWalle2010,FreysoldtGrabowskiHickelEtAl2014,BrobergMedasaniZimmermannEtAl2016} which separates the electrostatic interactions into a short range (decaying to zero in a large supercell) and a long range ($\sim \frac{1}{\varepsilon r}$ beyond the supercell boundaries) component. The dielectric constants ($\varepsilon$) of the spinels  utilized to approximate the long range part of the electrostatic interactions, are reported in Table~S2.   For a given defect, the value of $E_{corr}$ within the Freysoldt scheme{\cite{FreysoldtNeugebauerWalle2009,FreysoldtNeugebauerWalle2010}} is determined by the convergence of the short range potential to a constant value with increasing supercell size (as seen in Figure~S1). Recently, Komsa \emph{et al.}{\cite{KomsaRantalaPasquarello2012}} determined the Freysoldt scheme to be more efficient than other schemes,{\cite{LeslieGillan1985,MakovPayne1995,CarloniBloechlParrinello1995,Schultz2000,LanyZunger2008,KomsaRantalaPasquarello2012}} in terms of the supercell size required to achieve convergence, and quantified the average error of the Freysoldt correction to be $\sim~0.09$~eV in a variety of systems.{\cite{KomsaRantalaPasquarello2012}}
Specifically for the MgA$_2$Z$_4$ spinels (A~=~Sc/In, Z~=~S/Se), we use a 2$\times$2$\times$2 supercell of the conventional cubic cell, which contains 256 anions.

Throughout the article the Kr\"{o}ger--Vink notation is employed to identify the type of point defects in the MgA$_2$Z$_4$ spinels, including chalcogenide vacancies (e.g., Vac$_{\rm Se}$), metal vacancies (Vac$_{\rm Mg}$, Vac$_{\rm Sc}$), chalcogenide anti-sites (Se$_{\rm Mg}$, Se$_{\rm Sc}$) and metal anti-sites (Mg$_{\rm Sc}$, Sc$_{\rm Mg}$). For example, Sc$^{\bullet}_{\rm Mg}$ identifies a positively charged anti-site defect, where a Sc$^{3+}$ atom replaces Mg$^{2+}$. Similarly, Vac$^{''}_{\rm Mg}$ and Mg$^{\times}_{\rm Sc}$ represent a double-negatively charged vacancy on a Mg site and a neutral Mg anti-site on Sc, respectively.

\subsection{Charge neutrality} 
Point defects can be neutral or charged specie. An example is shown in Figure~S2, where the donor defect is positively charged for ($q = 1$) $E_{\rm Fermi} <~\epsilon(+/0)$, while for $E_{\rm Fermi} >~\epsilon(+/0)$ the donor defect is neutral ($q = 0$).  Thus, $\epsilon(+/0)$ is the thermodynamic defect transition level where two different charge states of a defect have the same  $E_f$. The availability of electrons  is set by the equilibrium Fermi level $E_{\rm Fermi}^{eq}$, and the defect transition level sets the $E_{Fermi}^{eq}$ within the band gap and in turn, the electronic conductivity of the structure.

\begin{figure*}[t]
\includegraphics[scale=0.350]{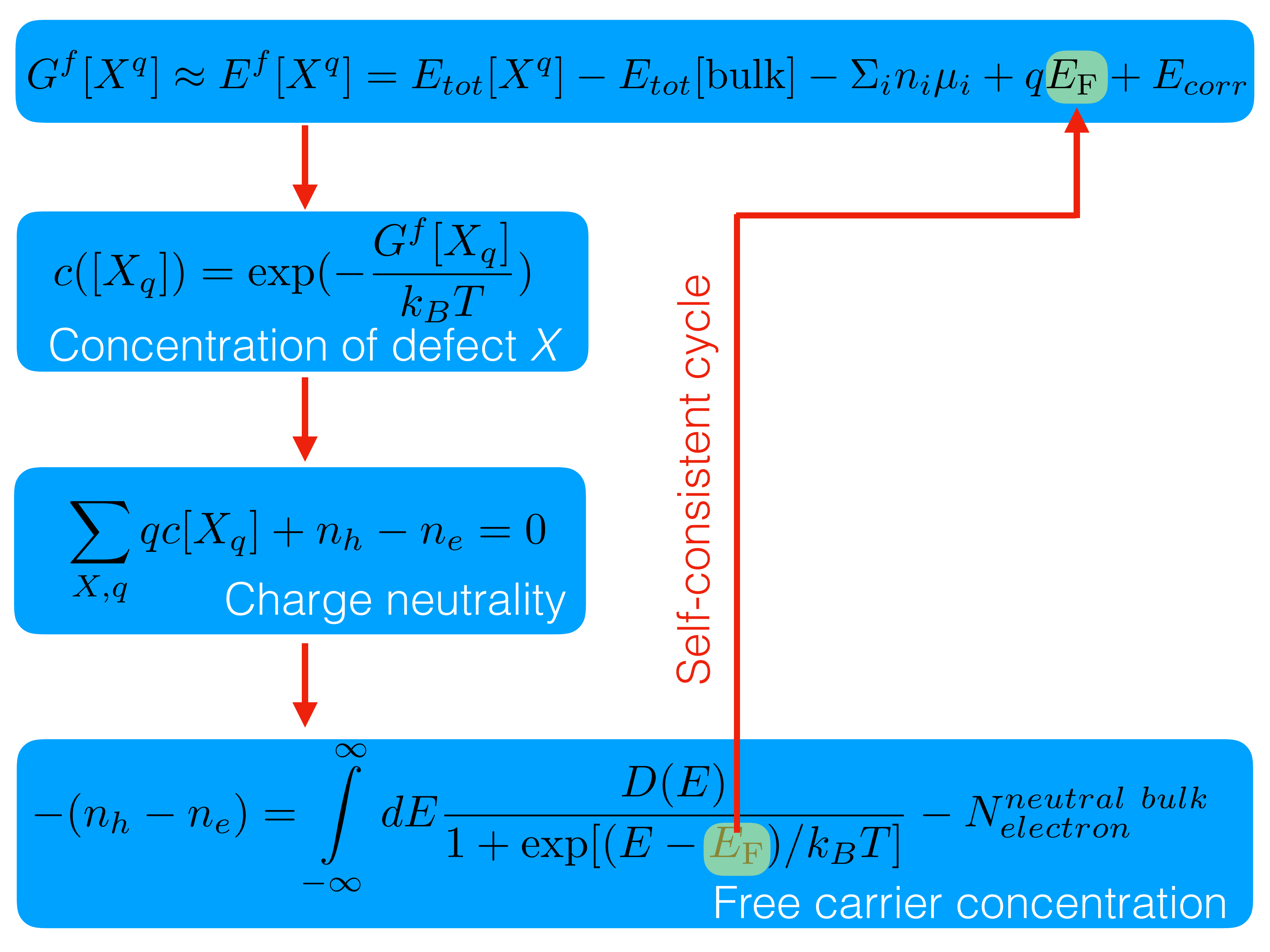}
\caption{
\label{fig:self}  Self-consistent search of the equilibrium Fermi level ($E_{F}$), defect ($c[X^q]$) and free carrier ($n_h$ and $n_e$) concentrations at a given temperature (T). The chemical potentials ($\mu_i$) are required as input parameters.}
\end{figure*}

When multiple defects and charge states are present in a structure, estimation of $E_{Fermi}^{eq}$ requires a self-consistent search (as explained  in Figure~{\ref{fig:self}}) by enforcing charge neutrality of the material, corresponding to $\sum_{X, q} qc[X^q] + n_h - n_e = 0$. $n_h$ ($n_e$) is the hole (electron) concentration, obtained by integrating the density of states (DOS, $D(E)$ in Figure~{\ref{fig:self}}) at a given Fermi level ($E_{F}$) in the structure. $N_{electron}^{neutral~bulk}$ in Figure~{\ref{fig:self}} is the total number of electrons in the neutral bulk cell. $c[X^q]$ is the concentration of defect $X^q$, stemming from the Gibbs energy of defect formation, $G_f[X^q]$~$\approx$~$E_f[X^q]$, as $c[X^q]\approx \exp(-E_f[X^q]/k_BT)$. The resulting $E_{Fermi}^{eq}$ and defect concentrations correspond to thermodynamic equilibrium as a function of temperature. Note that in all the materials considered in this work (Section~{\ref{sec:results}}), we list a few defects as ``dominant'' owing to their low formation energies at $E_{Fermi}^{eq}$.

%``

Materials that are normally synthesized at a high temperature ($\sim$~1273~K, for example\cite{CanepaBoGopalakrishnanEtAl2016}) and rapidly cooled to room temperature, may have their high temperature intrinsic defect concentrations ``frozen-in'' (or quenched) at room temperature, while the free carrier concentration ($n_h - n_e$) changes with temperature, given the fixed defect concentration. A change in intrinsic defect concentration will require significant atomic diffusion, which is likely to be kinetically limited at low temperatures. Hence, for calculating defect concentrations and the Fermi level, we have considered two scenarios within the constraint of charge neutrality, ($i$) defect concentrations, equilibrium Fermi level ($E_{Fermi}^{eq}$) and free electron/hole concentrations ($c[e/h]^{eq}$) are self-consistently calculated at 300~K corresponding to equilibrium conditions (as in Figure~{\ref{fig:self}}), ($ii$) defect concentrations are quenched from a higher synthesis temperature while the resulting Fermi level ($E_{Fermi}^{frozen}$) and free carrier concentrations ($c[e/h]^{frozen}$) are computed at 300~K.   When quenched or frozen conditions are assumed, the defect concentrations are calculated self-consistently at a higher quench temperature (i.e. 1273~K with the procedure in Figure~{\ref{fig:self}}), and are not allowed to change when the Fermi level and the free-carrier concentrations are re-calculated at 300~K. Since defect concentrations increase with increasing temperatures, the frozen approximation can quantify the possible deviations away from equilibrium in both defect and free carrier concentrations at 300~K.

\subsection{Computational Details}
The total energies in Eq.~\ref{eq:defecformation}  are obtained with Density Functional Theory\cite{HohenbergKohn1964,KohnSham1965} (DFT) using the Perdew Burke Ernzerhof (PBE) functional within the spin-polarized Generalized Gradient Approximation,\cite{PerdewBurkeErnzerhof1996}  as implemented in the VASP code.\cite{KresseHafner1993,KresseFurthmueller1996} Projector Augmented Wave theory\cite{Bloechl1994,KresseJoubert1999} and a plane-wave basis set with a cutoff of 520~eV are used to describe the crystalline wave-functions, which are subsequently sampled on a dense (minimum of 1000 $k$-points per atom in reciprocal space) $\Gamma$-centered $k$-point mesh.\cite{MonkhorstPack1976}  The Python Materials Genomics (pymatgen) \cite{OngRichardsJainEtAl2013} and the Python Charged Defect Toolkit (PyCDT){\cite{BrobergMedasaniZimmermannEtAl2016}} libraries are leveraged for input preparation and data analysis. For calculating the chemical potentials of the various species involved in the defect calculations, we utilize the Materials Project database\cite{JainOngHautierEtAl2013} in addition to our own calculations.%, to build the 0~K phase diagrams.

In order to sample the large chemical space of defects, we use the computationally inexpensive semi-local PBE exchange-correlation functional, especially because the  nature of the valence and conduction bands of the spinels considered, which are populated by the chalcogen (S/Se) and the metal (In/Sc) states respectively,  do not change with a higher level of theory, such as HSE06\cite{HeydScuseriaErnzerhof2003,HeydScuseriaErnzerhof2006} (see Figure~S3\cite{LejaeghereBihlmayerBjorkmanEtAl2016} in SI). 
To confirm this hypothesis we performed calculations of the low-lying defects in MgIn$_2$S$_4$ (see Section~{\ref{sec:results_MgIn2S4}}) utilizing both GGA and HSE06, and the comparison is detailed in Section~S9 of the SI.
In general, PBE is known to underestimate the band gap in most solids (by at least 30\%{\cite{Wang1983}}) when compared to HSE06 (Figure~S3).  For example, the PBE-computed direct band gaps are $\sim$~1.77~eV, $\sim$~1.56~eV, and $\sim$~1.09~eV for the MgIn$_2$S$_4$, MgSc$_2$S$_4$ and MgSc$_2$Se$_4$, respectively, while the magnitude of the gaps increase with HSE06,\cite{ChevrierOngArmientoEtAl2010} $\sim$~2.82~eV in MgIn$_2$S$_4$, $\sim$~2.63~eV in MgSc$_2$S$_4$, and $\sim$~2.03~eV in MgSc$_2$Se$_4$ (Figure~S3). Note that the band gaps decrease while moving down the chalcogenide group (i.e., S~$\rightarrow$~Se), under both PBE and HSE06 calculations.

 %%%%%%%%%%%%%%%%%%%%%%%%%%%%%%%%%%%%%%%%%%%%%%%%%%%%%%%%%%%%%%%%%%%%%%%%

 %%%%%%%%%%%%%%%%%%%%%%%%%%%%%%%%%%%%%%%%%%%%%%%%%%%%%%%%%%%%%%%%%%%%%%%%
 \section{MgA$_2$Z$_4$ structure and phase diagram}
\begin{figure*}[t]
\includegraphics[scale=0.20]{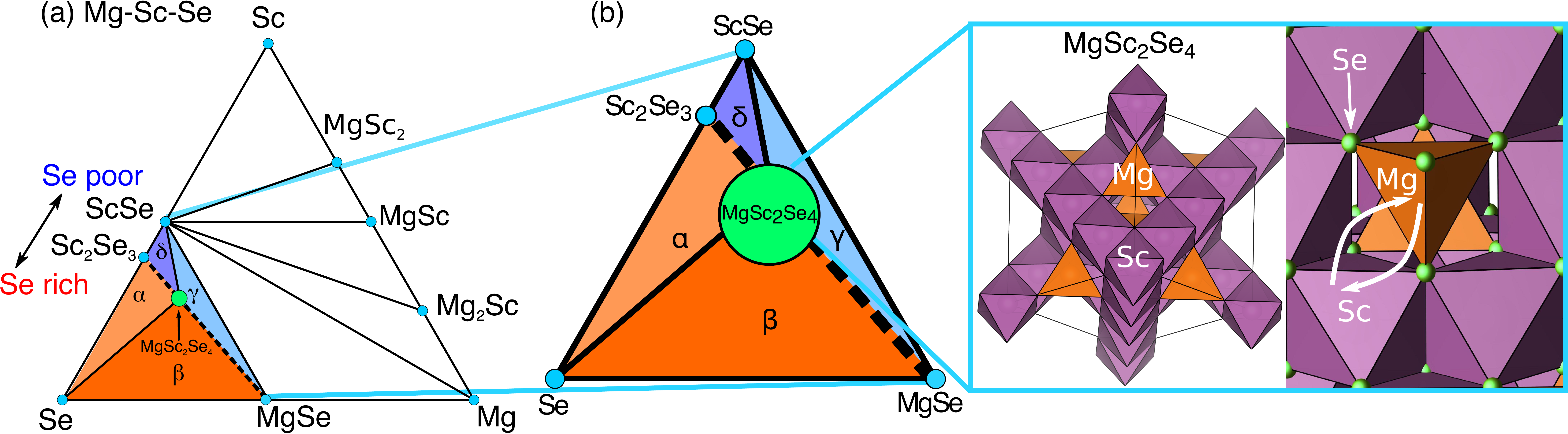}
\caption{
\label{fig:pdstructure} (a) Ternary Mg-Sc-Se phase-diagram at 0~K computed from DFT data combined with Materials Project,\cite{JainOngHautierEtAl2013} with (b) displaying a zoom-in of the concentration range of interest. (c) Crystal structure of a normal spinel, such as MgSc$_2$Se$_4$ identified in the phase diagrams of panels (a) and (b). The right fragment in (c) shows the scenario of spinel inversion (white arrows Mg~$\leftrightarrow$~Sc) in MgSc$_2$Se$_4$, leading to anti-site Mg$_{\rm Sc}$ and Sc$_{\rm Mg}$ defects. Similar ternary phase diagrams for  MgIn$_2$S$_4$  and MgSc$_2$S$_4$ are shown in Figure~S4.}
\end{figure*}

  %%%%%%%%%%%%%%%%%%%%%%%%%%%%%%%%%%%%%%%%%%%%%%%%%%%%%%%%%%%%%%%%%%%%%%%%
 The spinel structure  MgA$_2$Z$_4$ (with A = In or Sc and Z = S or Se), crystallizes with the anions in the face centered cubic (FCC) packing (space group: $Fd\bar{3}m$). In ``normal" spinel structures, the higher valent cations (A~=~In$^{3+}$ or Sc$^{3+}$),  occupy octahedral (oct) sites  $16d$, as shown by the purple polyhedra in Figure~\ref{fig:pdstructure}c, and the Mg$^{2+}$ occupy the tetrahedral (tet) $8a$ sites (orange polyhedra).  Few spinels, such as MgIn$_2$S$_4$, can also exhibit ``inversion'', as experimentally observed by Gastaldi~\emph{et al.},\cite{GastaldiLapiccirella1979} where a fraction of Mg$^{2+}$ ions in the $8a$ exchange sites with the In$^{3+}$ in $16d$. 

%`
  
The ternary 0~K phase-diagrams of Figures~\ref{fig:pdstructure}a and b depict four phases: Se, MgSe, ScSe and  Sc$_2$Se$_3$ that can be in thermodynamic equilibrium with the ternary MgSc$_2$Se$_4$ spinel, at different atomic chemical potentials ($\mu_{\rm Se}$ and $\mu_{\rm Mg}$). Equivalent phase diagrams have been constructed for the  Mg-In-S  and Mg-Sc-S systems and are presented in Figure~S4a and S4b of the SI. The four different facets of Figure~{\ref{fig:pdstructure}a and b, namely $\alpha$  MgSc$_2$Se$_4$-Se-Sc$_2$Se$_3$ (light orange), $\beta$ MgSc$_2$Se$_4$-Se-MgSe (dark orange), $\gamma$ MgSc$_2$Se$_4$-MgSe-ScSe (dark violet), and $\delta$ MgSc$_2$Se$_4$-ScSe-Sc$_2$Se$_3$ (light violet), define the possible limiting chemical potential values ($\mu _i$ of Eq.~\ref{eq:defecformation}) for intrinsic point defect formation, such as vacancies (e.g., Vac$_{\rm Mg}$) and anti-sites (e.g., Mg$_{\rm Sc}$). Subsequently, the  $\alpha$ and $\beta$ facets can be classified as ``Se rich'' domains, owing to elemental Se forming one of the bounding vertices of the respective facets, while $\gamma$ and $\delta$ are ``Se poor''. The dashed line in Figure~{\ref{fig:pdstructure}}a and b highlights the binary precursors, MgSe and Sc$_2$Se$_3$, which are used for the high-temperature synthesis ($\sim$~1200~$^{\circ}$C) of MgSc$_2$Se$_4$.\cite{PatrieFlahautDomage1964} Off-stoichiometry of MgSc$_2$Se$_4$, will place the thermodynamic equilibrium during synthesis into one of the four facets ($\alpha$ to $\delta$), which in turn can influence the  $E_f[X^q]$ and the defect concentrations.

\section{Native defects}
\label{sec:results}
\subsection{MgSc$_2$Se$_4$}
Figure~\ref{fig:MgScSedefects} shows the formation energies  $E_f[X^q]$ of intrinsic defects in MgSc$_2$Se$_4$ obtained  for the chemical potential in each of the four facets in the Mg-Sc-Se system, namely  MgSc$_2$Se$_4$-Se-Sc$_2$Se$_3$ $\alpha$ (Figure~\ref{fig:MgScSedefects}a), MgSc$_2$Se$_4$-Se-MgSe  $\beta$ (Figure~\ref{fig:MgScSedefects}b), MgSc$_2$Se$_4$-MgSe-ScSe $\gamma$ (Figure~\ref{fig:MgScSedefects}c), and MgSc$_2$Se$_4$-MgSe-Sc$_2$Se$_3$ $\delta$ (Figure~\ref{fig:MgScSedefects}d). The $y$-axis of each panel in Figure~{\ref{fig:MgScSedefects}} plots the defect energy against the $E_{Fermi}$ ($x$-axis) in MgSc$_2$Se$_4$. The absolute value of the Fermi energy is referenced  to the Valence Band Maximum (VBM) energy of the pristine MgSc$_2$Se$_4$ bulk. The zero of the $x$-axis is the VBM, with grey shaded regions being the valence ($E_{Fermi} < 0$) and the conduction bands ($E_{Fermi} > E_{gap} \sim$~1.09~eV), respectively. The  band gap spans the white area in all panels of Figure~\ref{fig:MgScSedefects}. In general, the defect levels with low formation energies in the band gap can considerably alter the intrinsic electronic conductivity of semiconductors and insulators, thus forming the region of interest in this analysis.

\begin{figure*}[t]
\includegraphics[scale=0.37]{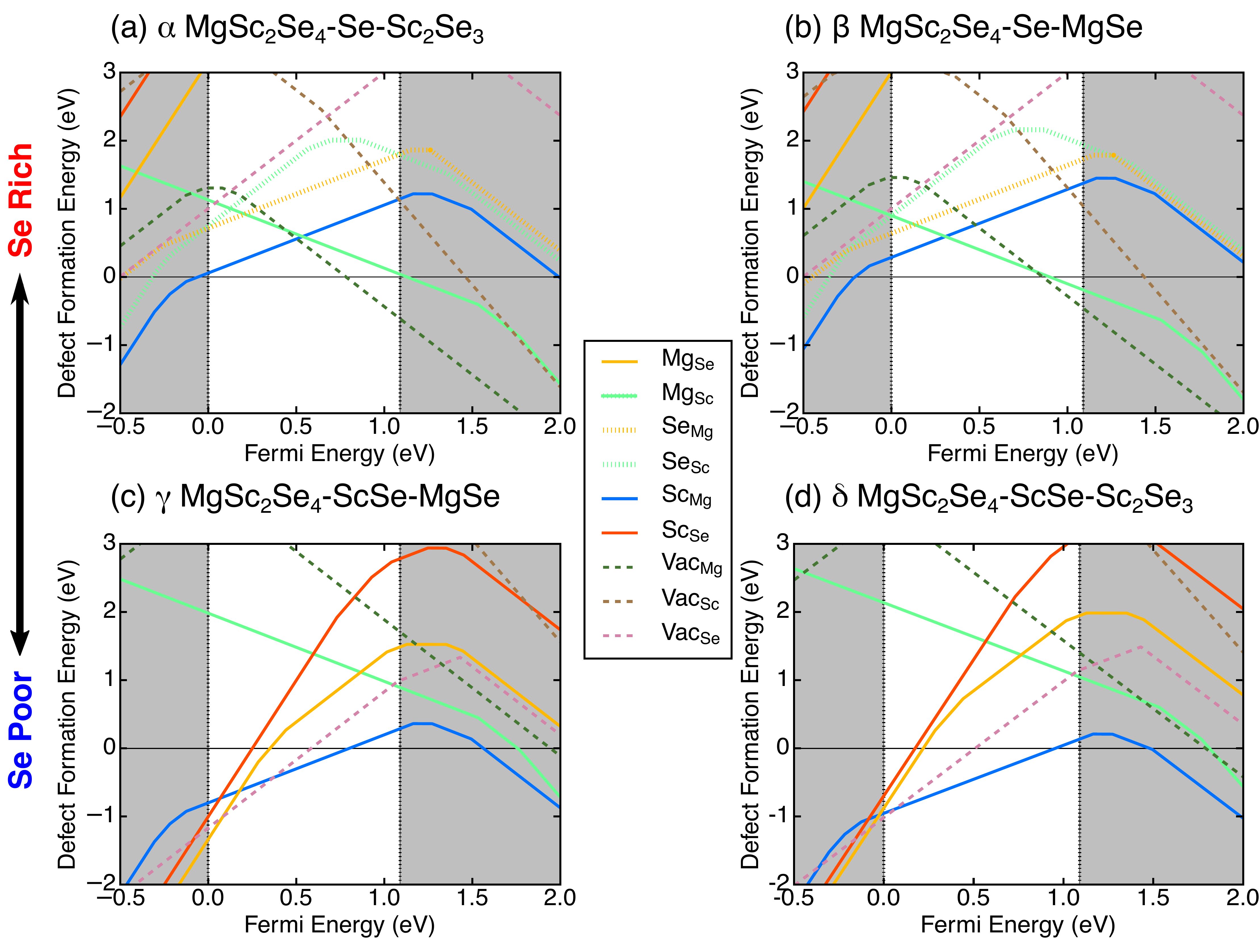}
\caption{
\label{fig:MgScSedefects} Defect energy $E_f[X^q]$ for intrinsic point defects (in Kr\"{o}ger--Vink notation) of MgSc$_2$Se$_4$ in four facets,  $\alpha$, $\beta$, $\gamma$ and $\delta$, of the Mg-Sc-Se phase diagram. Regions $\alpha$ and  $\beta$ are Se-rich, whereas $\gamma$ and $\delta$ Se-poor. The VBM is arbitrarily set to 0 eV and the white region spans the band-gap ($\sim$~1.09~eV). Vac in the legend and dashed lines indicate vacancy defects while solid lines correspond to anti-site defects.}
\end{figure*}

Facets $\alpha$ and $\beta$ (Figures~\ref{fig:MgScSedefects}a and b)  are Se-rich, and show qualitatively similar defect energetics.  For example, the defects with the lowest $E_f[X^q]$ are the  Sc$_{\rm Mg}$, Mg$_{\rm Sc}$ and Vac$_{\rm Mg}$ in both $\alpha$ and $\beta$. Sc$_2$Se$_3$ has been previously been detected as a prominent impurity in the synthesis of MgSc$_2$Se$_4$,\cite{CanepaBoGopalakrishnanEtAl2016} thus motivating the choice of  facet $\alpha$ (MgSc$_2$Se$_4$-Se-Sc$_2$Se$_3$) to characterize the Se-rich domain.  Similar  conclusions  are deduced by comparing the $\gamma$ and $\delta$ facets (Figure~{\ref{fig:MgScSedefects}}c and d),  with comparable  $E_f[X^q]$  for the low lying defects (e.g., Sc$_{\rm Mg}$), and only the $\gamma$ phase is considered further to analyze the Se-poor domain. Analogous behaviors are also observed for MgSc$_2$S$_4$ and MgIn$_2$S$_4$, showing similar trends for the S-rich and S-poor domains (Figure~S5 and S6 in the SI).

\subsubsection*{Se-rich domain, $\alpha$ MgSc$_2$Se$_4$-Se-Sc$_2$Se$_3$}
The dominant defects within the band gap of the Se-rich region are the charged Sc$^{\bullet}_{\rm Mg}$, Mg$^{'}_{\rm Sc}$ and Vac$^{''}_{\rm Mg}$ (dark blue, light green and dashed green lines, respectively, in Figure~\ref{fig:MgScSedefects}a).  A charged defect always exchanges its excess (deficient) charge with the electron reservoir of the structure, whose energy is given by the Fermi energy.  Thus, the $n$-type Sc$^{\bullet}_{\rm Mg}$ exchanges the excess valence electron from Sc with the Fermi level of MgSc$_2$Se$_4$. Analogous considerations extend to the $p$-type Mg$^{'}_{\rm Sc}$, where one electron is added to the anti-site from the Fermi level.

Given that the opposite charges of Sc$^{\bullet}_{\rm Mg}$, Mg$^{'}_{\rm Sc}$ and Vac$^{''}_{\rm Mg}$ can potentially charge-compensate each other leading to charge neutrality, the $E_{Fermi}^{eq}$ is nominally pinned at a Fermi energy where all three defects have similar $E_f$. Indeed, a self-consistent calculation of the $E_{Fermi}^{eq}$ at 300 K (i.e., assuming defect concentrations equilibrate at 300~K) leads to a $E_{Fermi}~=$0.46~eV (see Figures~\ref{fig:MgScSedefects}a and S5a in SI), with defect concentrations of 7.9$\times$10$^{11}$ cm$^{-3}$ for Sc$^{\bullet}_{\rm Mg}$,  2.4$\times$10$^{11}$ cm$^{-3}$ for Mg$^{'}_{\rm Sc}$ and 2.8$\times$10$^{11}$ cm$^{-3}$ for Vac$^{''}_{\rm Mg}$. Typically, defect contents above 10$^{15}$ cm$^{-3}$  are detectable via experiments, such as electron paramagnetic resonance.\cite{Weil1984,Livraghi2006,FreysoldtGrabowskiHickelEtAl2014} Alternatively, defect concentrations can be expressed in units of per atom or per formula unit. For example, a defect concentration of 10$^{15}$~cm$^{-3}$ in MgSc$_2$Se$_4$ corresponds to $\sim$~2.4$\times 10^{-8}$/atom and 1.7$\times 10^{-7}$/(formula unit), respectively. For the $\alpha$-facet of MgSc$_2$Se$_4$ at 300 K, the Fermi level is ``deep'' within the band gap, which will lead to low electronic (or hole) conductivity since large thermal energies ($\gg k_B T$) will be required to ionize free electrons (holes) from the $E_{Fermi}^{eq}$ into the conduction (valence) band. Qualitatively similar conclusions can be drawn from an analysis of the defects in the $\beta$ facet (Figure~{\ref{fig:MgScSedefects}}b and Figure~S7a). 

However, when defect concentrations are frozen-in from a higher temperature ($\sim$~1273~K used for MgSc$_2$Se$_4$ synthesis{\cite{CanepaBoGopalakrishnanEtAl2016}}), the $E_{Fermi}^{frozen}$ at 300~K drops below the VBM ($\sim$~--0.10~eV) indicating that the material becomes a $p$-type conductor. Thus, significant hole conductivity can be expected under frozen defect conditions, with free hole concentration of $\sim$~2.6$\times$10$^{18}$~cm$^{-3}$  ($\sim$~0.0001 per lattice site), which is beyond un-doped semiconductor levels ($\sim~10^{10}$~cm$^{-3}$ in SI) but below metallic levels ($\sim$~1 charge carrier per lattice site). As the temperature at which the defect concentrations are quenched decreases, the $E_{Fermi}^{frozen}$ recovers beyond the VBM and reaches $\sim$~0.02~eV at 800~K (Figure~S7a), indicating the importance of slow cooling conditions to reduce hole conductivity during the synthesis of MgSc$_2$Se$_4$. 

\subsubsection*{Se-poor domain, $\gamma$ MgSc$_2$Se$_4$-ScSe-MgSe}
The Se-poor region (Figure~\ref{fig:MgScSedefects}c) is dominated by $n$-type defects, such as Sc$^{\bullet}_{\rm Mg}$ (dark blue), Vac$^{\bullet \bullet}_{\rm Se}$ (dashed red), and Mg$^{\bullet \bullet \bullet}_{\rm Se}$ (orange). Although the formation energies of a few defects are negative across the band gap, as in the case of  Sc$^{\bullet}_{\rm Mg}$ for $E_{Fermi} <$~0.7~eV (Figure~\ref{fig:MgScSedefects}c), the spontaneous formation of such charged defects is constrained by the condition of charge neutrality in MgSc$_2$Se$_4$.

The self-consistent equilibrium Fermi level ($\sim$~1.08~eV at 300~K) for the Se-poor region is mainly set by the Sc$^{\bullet}_{\rm Mg}$ defect. However for temperatures above 300 K, the $E_{Fermi}^{eq}$ exceeds the Conduction Band Minimum (CBM, $\sim$~1.1~eV, Figure~S7a), suggesting the occurrence of spontaneous electronic conductivity when the spinel is synthesized under Se-poor conditions. Furthermore, when defect concentrations are frozen-in from $\sim$~1273~K, $E_{Fermi}^{frozen}$ is well above the CBM ($\sim$~1.4~eV) at 300~K, suggesting that fast cooling during synthesis will likely increase the electronic conductivity. Similar conclusions can be extended by evaluating the defect energies in the $\delta$ facet (Figure~\ref{fig:MgScSedefects}d, and Figure~S7a in the SI), where the equilibrium Fermi level is beyond the CBM even at 300~K, suggesting that preventing intrinsic electronic conductivity in MgSc$_2$Se$_4$ in Se-poor conditions may be challenging. 
 
\subsection{MgIn$_2$S$_4$}
\label{sec:results_MgIn2S4}
Figure~\ref{fig:MgInSdefects} plots the defect formation energies for the S-rich ($\alpha$) and S-poor ($\gamma$) domains as a function of Fermi energy in the MgIn$_2$S$_4$ spinel. Analogous to MgSc$_2$Se$_4$ (Figure~{\ref{fig:MgScSedefects}}a),  the stable defects in the S-rich domain of MgIn$_2$S$_4$ (Figure~{\ref{fig:MgInSdefects}}a), include the $n$-type In$^{\bullet}_{\rm Mg}$ (solid blue), and the $p$-type Mg$^{'}_{\rm In}$ (light green) and Vac$^{''}_{\rm Mg}$ (dashed green). For the $\alpha$-facet (S-rich domain), the resulting $E_{Fermi}^{eq}$ is $\sim$~0.88~eV at 300~K, corresponding to a free  hole concentration of $\sim$~6.46$\times$10$^4$~cm$^{-3}$. Other notable defects, such as Vac$^{'''}_{\rm In}$, Vac$^{\bullet \bullet}_{\rm S}$ and S$^{'}_{\rm In}$ are not expected to play a dominant role in MgIn$_2$S$_4$. 

\begin{figure}[!ht]
\includegraphics[scale=0.40]{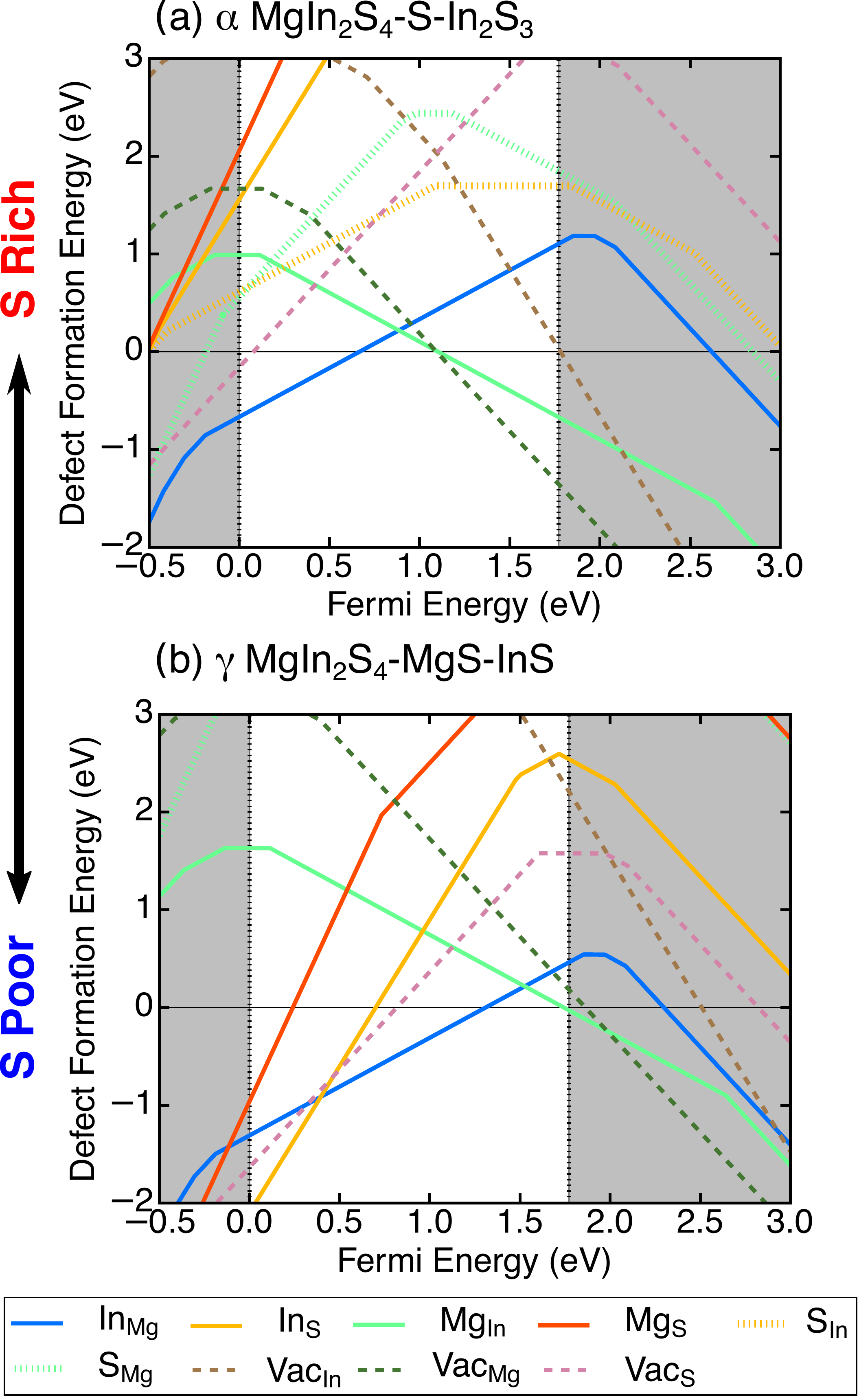}
\caption{
\label{fig:MgInSdefects} (Colors online)
 $E_f[X^q]$ for intrinsic point defects (in Kr\"{o}ger--Vink notation) of MgIn$_2$S$_4$ for two regions of the phase diagram ($\alpha$ and $\gamma$, refer to Figure~\ref{fig:MgScSedefects}). Region $\alpha$ is S-rich, whereas $\gamma$ is S-poor. The VBM is set to 0 eV and the white region is the band-gap ($\sim$~1.77~eV). Vac in legend and dashed lines indicate vacancy defects while solid lines correspond to anti-site defects.  }
\end{figure}

Since the equilibrium Fermi energy is pinned near the middle of the band-gap ($\sim$~0.88~eV Figure~{\ref{fig:MgInSdefects}}a) by self-compensating charged defects, the material will exhibit low electronic conductivity under equilibrium S-rich conditions. Nevertheless, the equilibrium defect concentrations are significant, $\sim$ 4.9$\times$10$^{17}$~cm$^{-3}$ for In$^{\bullet}_{\rm Mg}$, 4.9$\times$10$^{17}$~cm$^{-3}$ Mg$^{'}_{\rm In}$ and 7.8$\times$10$^{13}$~cm$^{-3}$ Vac$^{''}_{\rm Mg}$. Such high  concentrations of anti-site In$^{\bullet}_{\rm Mg}$ and Mg$^{'}_{\rm In}$ defects indicate that the spinel undergoes a high degree of ``inversion" ---Mg and In exchanging their lattice sites{\cite{Wakaki1980}}--- besides exhibiting significant Mg vacancies during S-rich synthesis conditions.

The analysis of the defect formation energies in the S-poor facet ($\gamma$, Figure~\ref{fig:MgInSdefects}b) suggests that the Vac$^\bullet _{\rm S}$ and In$^{\bullet \bullet} _{\rm S}$ defects can influence $E_{Fermi}^{eq}$, apart from the In$^{\bullet}_{\rm Mg}$, Mg$^{'} _{\rm In}$ and Vac$^{''}_{\rm Mg}$. Interestingly, the self-consistent Fermi level of the $\gamma$ facet is $\sim$~1.53~eV, which corresponds approximatively to charge-compensation between the In$^{\bullet}_{\rm Mg}$ and Mg$^{'}_{\rm In}$ defects. Furthermore, the equilibrium defect concentrations ($\sim$~4.9$\times$10$^{17}$ for both In$^{\bullet}_{\rm Mg}$ and Mg$^{'}_{\rm In}$, respectively) estimated under S-poor conditions compare well with a S-rich environment, indicating that the MgIn$_2$S$_4$ is expected to undergo a substantial degree of spinel inversion irrespective of synthesis conditions, in agreement with experimental observations.{\cite{Wakaki1980}} 

Similar to MgSc$_2$Se$_4$, cooling rates during synthesis are expected to play a major role in determining the intrinsic hole/electronic conductivity in the In-spinel (Figure~S7b). For example, under frozen-in defect concentrations from $\sim$~1273~K, the $E_{Fermi}^{frozen}$ and $c[e/h]^{frozen}$ are $\sim$~0.10~eV, 1.17$\times$10$^{18}$ (free holes) for S-rich and $\sim$~1.80~eV ($>$~CBM), $1.12\times$10$^{19}$ (free electrons) for S-poor, respectively.  MgIn$_2$S$_4$ is expected to exhibit $p$-type and $n$-type conductivity in S-rich and S-poor conditions, respectively, under quenched defect concentrations. Thus, the protocols to synthesize MgIn$_2$S$_4$ requires careful tuning to allow for slow cooling and S-rich conditions.

\subsection{MgSc$_2$S$_4$}
Figure~\ref{fig:MgScSdefects} shows the formation energies of  native  defects of MgSc$_2$S$_4$ in both S-rich ($\alpha$ facet) and S-poor ($\gamma$ facet) domains. The defect energetics in MgSc$_2$S$_4$ are similar to the Se-spinel, in the anion-rich domain (Figure~{\ref{fig:MgScSdefects}}a), with the dominant defects being Sc$^{\bullet}_{\rm Mg}$, Mg$^{'}_{\rm Sc}$, and Vac$^{''}_{\rm Mg}$. The $E_{Fermi}^{eq}$, calculated self-consistently at 300~K is $\sim$~0.4~eV and roughly corresponds to the self-compensation of  Vac$^{''}_{\rm Mg}$ and Mg$^{'}_{\rm Sc}$ with Sc$^{\bullet}_{\rm Mg}$. Therefore, when MgSc$_2$S$_4$ is prepared under S-rich conditions it should exhibit a small degree of spinel inversion (Sc$^{\bullet}_{\rm Mg}$ 1.7$\times$10$^{11}$ cm$^{-3}$ and Mg$^{'}_{\rm Sc}$ $\sim$3.7$\times$10$^{11}$ cm$^{-3}$), and low hole conductivity ({$c[h]^{eq} \sim$ 2.01$\times$$10^{11}$~cm$^{-3}$). Also, under frozen-in defect conditions (from 1273~K), MgSc$_2$S$_4$ becomes a spontaneous $p$-type conductor similar to MgSc$_2$Se$_4$, with $E_{Fermi}^{frozen}$ drifting below the VBM ($\sim-0.06$~eV) resulting in a larger $c[h]^{frozen} \sim 1.18$$\times$$10^{19}$~cm$^{-3}$. 

\begin{figure}[!ht]
\includegraphics[scale=0.40]{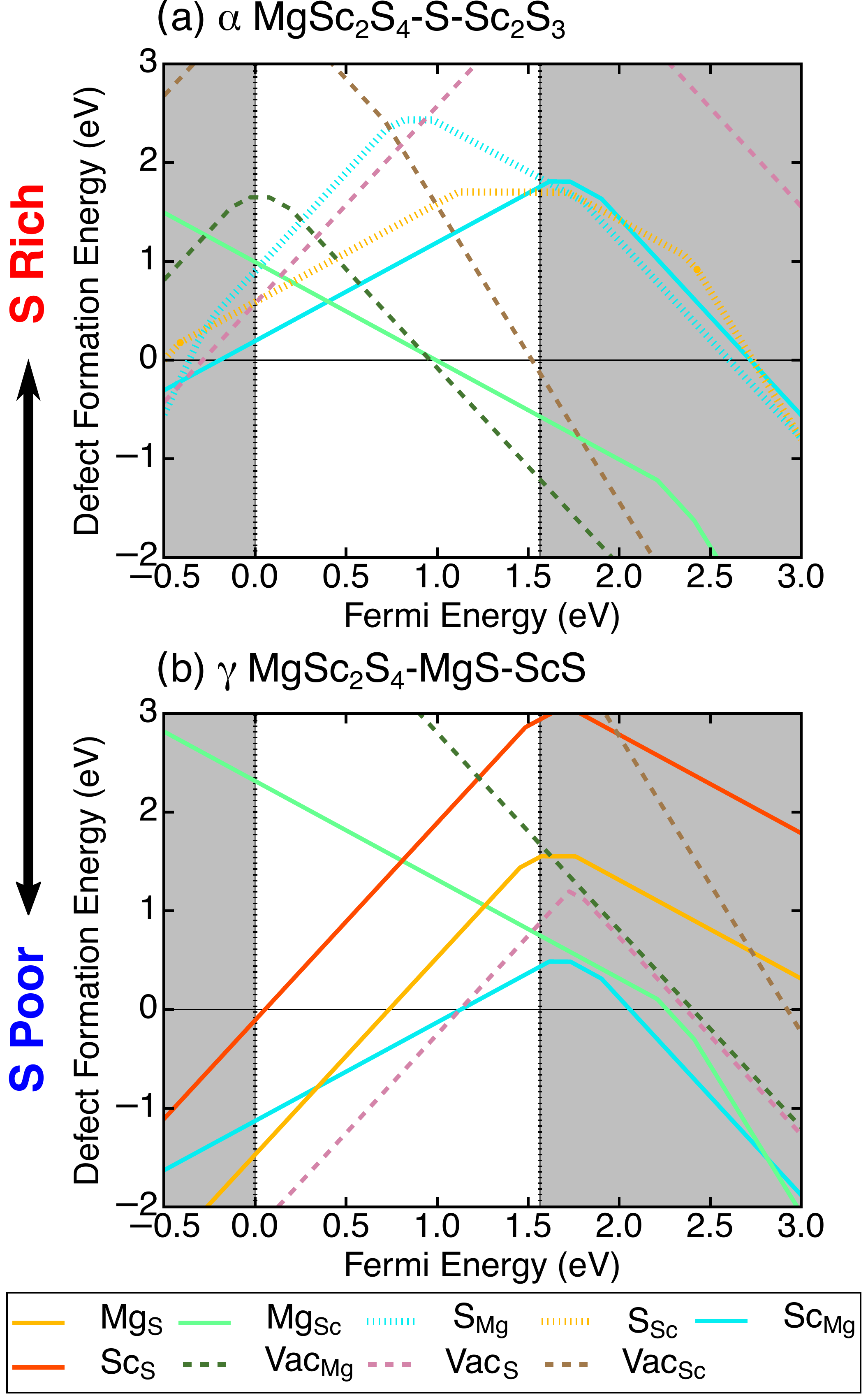}
\caption{
\label{fig:MgScSdefects} $E_f[X^q]$ for intrinsic point defects (in Kr\"{o}ger--Vink notation) of MgSc$_2$S$_4$ for two regions of the phase diagram ($\alpha$ and $\gamma$, refer to Figure~\ref{fig:MgScSedefects}). Region $\alpha$ is S-rich, whereas $\gamma$ is S-poor. The VBM is set to 0 eV and the white region is the band-gap ($\sim$~1.56~eV). Vac in legend and dashed lines indicate vacancy defects and  solid lines  indicate anti-site  defects.  }
\end{figure}

The dominant point defects in the S-poor region ($\gamma$, Figure~{\ref{fig:MgScSdefects}}b) are Mg$^{\bullet \bullet}_{\rm S}$,  Sc$^{\bullet}_{\rm Mg}$, Vac$^{\bullet \bullet}_{\rm S}$ and Mg$^{'}_{\rm Sc}$, with Vac$^{\bullet \bullet}_{\rm S}$ displaying the lowest $E_f[X^q]$ across the band gap, up to $E_{Fermi}$ $\sim~1.5$~eV. In the $\gamma$ (and $\delta$, Figure~S7c) facet, the equilibrium Fermi level at 300~K is beyond the CBM, indicating spontaneous electronic conductivity under S-poor conditions. Under quenched defect conditions (from 1273~K), $E_{Fermi}^{frozen}$ is found to be deeper into the conduction band ($\sim$~1.80~eV) compared to equilibrium at 300~K (Figure~S7c). Indeed, $c[e]^{eq}$ is estimated to be $\sim$1.8$\times 10^{15}$~cm$^{-3}$ under equilibrium at 300~K, while the concentration increases by nearly 5 orders of magnitude ($c[e]^{frozen} \sim  7.8$$\times$$10^{19}$~cm$^{-3}$)  under quenched conditions. Thus, suppressing intrinsic electronic conductivity in MgSc$_2$S$_4$ under S-poor conditions represents a major challenge.

\section{Extrinsic defects in MgSc$_2$Se$_4$}
High ionic conductivity in materials is often achieved if the concentration of mobile vacancies is increased. One strategy commonly adopted to increase ionic conductivity in solid electrolytes is extrinsic doping, specifically doping the anion sub-lattice.{\cite{Miara2015,Janek2016,Kato2016}} Nominally, the selection of an extrinsic dopant follows the rule of thumb of finding similar-sized cations (anions) for aliovalent substitution in the lattice. In addition, it is desirable that the substituting element is not redox-active, which will minimize the occurrence of redox side-reactions in ionic conductors.

In the case of the spinel Mg-conductors discussed in this work, the electronic conductivity primarily arises from anti-site defects, such as Sc$^{\bullet}_{\rm Mg}$ and In$^{\bullet}_{\rm Mg}$, pushing the equilibrium Fermi level close (or beyond) the CBM level at 300~K.  A pathway to curb the formation of anti-site defects is doping the metal site (Sc or In) with cations that are less likely to promote spinel inversion. For example, metal ions with higher oxidations states (such as Zr$^{4+}$ and Nb$^{5+}$) or those manifesting a stronger octahedral site preference than Mg$^{2+}$, are less likely to occupy the tetrahedral spinel sites.\cite{Pauling1927} Thus, cation doping on the Sc site can inhibit the formation of anti-site Sc$_{\rm Mg}$ defects in MgSc$_2$Se$_4$ (and MgSc$_2$S$_4$).

Figure~{\ref{fig:extrinsics}} plots the defect formation energies for extrinsic doping of several \st{non-redox} tetravalent (Ce, Ge, Sn, Pb, Ti and Zr, solid lines in Figure~{\ref{fig:extrinsics}}a), and pentavalent (As, Bi, Na and Ta, dashed lines) cations on Sc in MgSc$_2$Se$_4$, as well as anion doping on Se$^{2-}$ with monovalent anions (Cl$^-$, Br$^-$ and I$^-$, Figure~{\ref{fig:extrinsics}}b). Because of the Se-poor synthesis conditions normally encountered, we restrict the analysis only to the $\gamma$ facet (MgSc$_2$Se$_4$-MgSe-ScSe), while calculations of the $\alpha$, $\beta$ and $\delta$ facets of MgSc$_2$Se$_4$ are discussed in Figures~S7 and S8 of the SI. Note that the chemical potential of each extrinsic dopant in Eq.~\ref{eq:defecformation} is set by the most stable phase in the Mg-Sc-Se-[extrinsic dopant] phase diagram that is in equilibrium with MgSc$_2$Se$_4$, MgSe, and ScSe (accessed via the Materials Project\cite{JainOngHautierEtAl2013}). For example, in the case of Cl$^{-}$ doping on Se$^{2-}$, $\mu_{\rm Cl}$ is determined by the facet MgSc$_2$Se$_4$-MgSe-ScSe-MgCl$_2$.

\begin{figure}[!ht]
\includegraphics[scale=0.4]{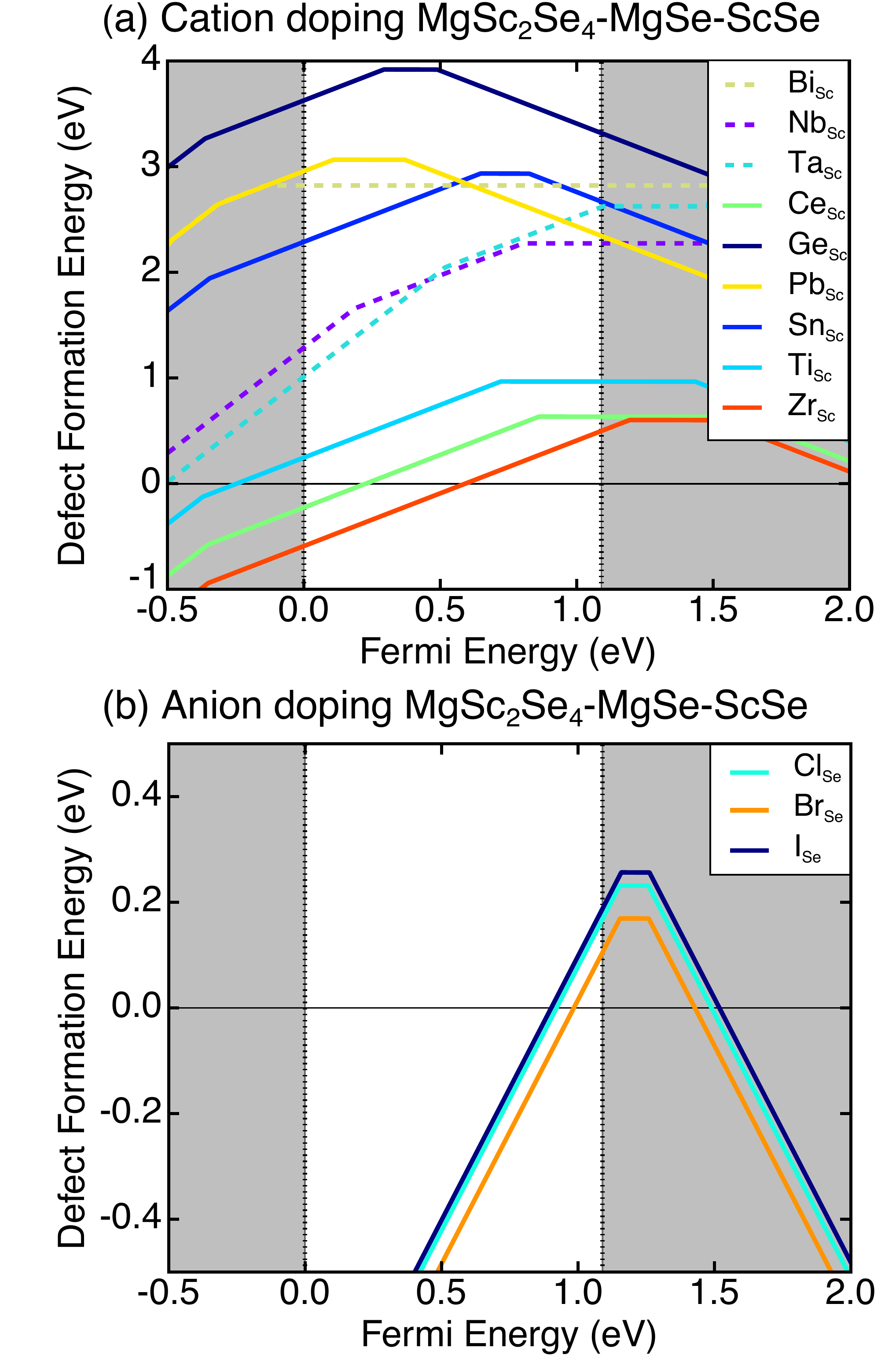}
\caption{
\label{fig:extrinsics} $E_f[X^q]$ for extrinsic cation (a) and anion (b) substitution in MgSc$_2$Se$_4$ under Se-poor conditions ($\gamma$-facet). In panel (a), tetravalent and pentavalent cations are shown by solid and dashed lines, respectively. The VBM is set to 0 eV and the white region is the band-gap ($\sim$~1.09~eV). $E_f[X^q]$ is not shown for As$_{\rm Sc}$ substitution, since the values are above 4 eV through the Fermi energy range considered.}
\end{figure}

Data in Figure~\ref{fig:extrinsics} suggests that extrinsic doping of several cations, such as Bi, Nb, Ta, Ge, Sn and Pb on Sc in MgSc$_2$Se$_4$ is highly unfavourable (with $E_f[X^q]$ $\geq$ 1~eV). In contrast, halogen doping on Se (Figure~{\ref{fig:extrinsics}}b), Ce (solid green line in Figure~\ref{fig:extrinsics}a), Ti (solid light blue), and Zr (solid red line) doping on Sc appear favorable. Specifically, halogen substitution on Se and Zr$^{\bullet}_{\rm Sc}$ show a negative formation energy over a wide portion of the band-gap. 

Since the behavior of halogen doping (and Zr$^{\bullet}_{\rm Sc}$) is similar to intrinsic $n$-type Sc$^{\bullet}_{\rm Mg}$ (Figure~{\ref{fig:MgScSedefects}}c) anti-sites, anion (and Zr) doping in MgSc$_2$Se$_4$ may not be beneficial since the $E_{Fermi}^{eq}$ is likely to be pushed into the conduction band at 300~K. However, $n$-type Ce$^{\bullet}_{\rm Sc}$ and Ti$^{\bullet}_{\rm Sc}$ show fairly deep donor transition levels away from the CBM (Figure~{\ref{fig:extrinsics}}a). Indeed, $E_{Fermi}^{eq}$ at 300~K for the Ce$^{\bullet}_{\rm Sc}$ and Ti$^{\bullet}_{\rm Sc}$, without considering the presence of intrinsic anti-sites, are $\sim$~0.82~eV and 0.58~eV, respectively. Thus, Ce and Ti doping should not increase the electronic conductivity of MgSc$_2$Se$_4$, though their effect on the Sc$_{\rm Mg}$ formation energies requires more investigation.

\section{Discussion}
Using first-principles defect energy calculations, we analyzed the defect chemistry in chalcogenide Mg spinels, namely MgSc$_2$Se$_4$, MgSc$_2$S$_4$ and MgIn$_2$S$_4$, and have summarized the Fermi energies and free-carrier concentrations in Table~{\ref{tb:defects}} (defect concentrations are also tabulated in Table~S1) for a representative anion-rich and anion-poor equilibrium. Under all conditions, anti-sites (Mg$_{\rm \{Sc/In\}}$ and \{Sc/In\}$_{\rm Mg}$) and Mg-vacancies are the dominant defects, while anion vacancies only show up for MgSc$_2$S$_4$ under S-poor conditions.
\begin{table}[!h]
\caption{\label{tb:defects} Defect energetics in the MgA$_2$Z$_4$ spinels (A = Sc, In, Z = S, Se), for both anion-rich ($\alpha$) and anion-poor ($\gamma$) conditions (facets). Self-consistent $E_{Fermi}^{eq}$  at 300~K (in eV)  and Fermi levels with quenched defect content (from 1273~K, $E_{Fermi}^{frozen}$), are indicated. $c[e/h]^{eq}$ and $c[e/h]^{frozen}$ (in cm$^{-3}$ at 300~K) are the free charge-carrier concentration in the self-consistent equilibrium and frozen defect scenarios, with $e$ and $h$ for electrons and holes. The charge of the dominant defect is indicated with respect to the charged state of the defect at $E_{Fermi}^{eq}$.}
\begin{tabular*}{\columnwidth}{@{\extracolsep{\fill}}llccccr@{}}
\hline\hline
Condition & Dominant defects  & Carrier & $E_{Fermi}^{eq}$   & $c[e/h]^{eq}$ & $E_{Fermi}^{frozen}$   & $c[e/h]^{frozen}$ \\
\hline \hline
\multicolumn{7}{c}{\textbf{MgSc$_2$Se$_4$ (E$_{gap}$ = 1.09~eV)}} \\
Se-rich ($\alpha$) & Sc$^{\bullet}_{\rm Mg}$, Mg$^{'}_{\rm Sc}$,Vac$^{''}_{\rm Mg}$ & $h^+$ &0.46 & 8.42$\times$$10^{8}$ &  --0.10 &2.58$\times$$10^{18}$ \\ 
Se-poor ($\gamma$) & Sc$^{\bullet}_{\rm Mg}$ & $e^-$ & 1.08 & 7.96$\times$$10^{15}$ &1.39 & 2.77$\times$$10^{19}$ \\  
\hline
\multicolumn{7}{c}{\textbf{MgIn$_2$S$_4$ (E$_{gap}$ = 1.77~eV)}} \\
S-rich ($\alpha$) & In$^{\bullet}_{\rm Mg}$, Mg$^{'}_{\rm In}$, Vac$^{''}_{\rm Mg}$ & $h^+$ & 0.88 & 6.46$\times$$10^{4}$ & 0.10 & 1.17$\times$$10^{18}$ \\ 
S-poor ($\gamma$) & In$^{\bullet}_{\rm Mg}$, Mg$^{'}_{\rm In}$, Vac$^{''}_{\rm Mg}$ & $e^-$ & 1.53 & 4.10$\times$$10^{14}$ & 1.80 & 1.12$\times$$10^{19}$ \\  
\hline
\multicolumn{7}{c}{\textbf{MgSc$_2$S$_4$ (E$_{gap}$ = 1.55~eV)}} \\
S-rich ($\alpha$) & Sc$^{\bullet}_{\rm Mg}$, Mg$^{'}_{\rm Sc}$, Vac$^{''}_{\rm Mg}$ & $h^+$ & 0.40 & 2.01$\times$$10^{11}$ & $-$0.06 & 1.18$\times$$10^{19}$ \\ 
S-poor ($\gamma$)  & Sc$^{\bullet}_{\rm Mg}$, Mg$^{'}_{\rm Sc}$, Vac$^{\bullet \bullet}_{\rm S}$ & $e^-$ & 1.48 & 1.81$\times$$10^{15}$ & 1.80 & 7.86$\times$$10^{19}$ \\  
\hline\hline
\end{tabular*}

\end{table}

\subsection{Anion-rich vs.\ anion-poor conditions}
All three spinels display markedly different defect energetics in the anion (S/Se)-rich and anion-poor domains, under equilibrium defect concentrations. In the case of anion-rich conditions ($\alpha$ facet), the spinels exhibit marginal $p$-type behavior with low carrier concentrations, due to the presence of charged anti-sites (\{Sc/In\}$^{\bullet}_{\rm Mg}$ and Mg$^{'}_{\rm \{Sc/In\}}$), and Vac$^{''}_{\rm Mg}$, which charge-compensate each other and pin the $E_{Fermi}^{eq}$ within the respective band gaps. Since the $E_{Fermi}^{eq}$ is far away from the VBM (or CBM), i.e., $\gg k_{B}T$, the hole (or electronic) conductivity is not expected to be significant (see Table~{\ref{tb:defects}}). Hence, the synthesis of the chalocogenide conductors in anion-rich environments should curtail, to a large extent, the undesired hole/electron conductivity for application as a Mg solid electrolyte. However, synthesis of the Se spinels  requires high temperatures ($>$~1000 $^{\circ}$C),\cite{PatrieFlahautDomage1964,GuittardSouleauFarsam1964,CanepaBoGopalakrishnanEtAl2016} at which elemental Se (b.p.~$\sim$~685~$^\circ$C) and S ($\sim$~444$~^\circ$C) vaporize and may lead to anion-poor conditions. One potential strategy to mitigate anion loss during synthesis is to use the respective stoichiometric binaries, such as MgSe and Sc$_2$Se$_3$ to form MgSc$_2$Se$_4$, at high temperature.\cite{PatrieFlahautDomage1964,GuittardSouleauFarsam1964}

Unlike anion-rich conditions, the dominant $n$-type Sc$^{\bullet}_{\rm Mg}$ anti-sites in the anion poor domain ($\gamma$ facet) push the $E_{Fermi}^{eq}$ beyond the CBM in both Sc-spinels, ensuring spontaneous electronic conductivity. We speculate that the low Sc$^{\bullet}_{\rm Mg}$ equilibrium concentration ($\sim$~7.96$\times 10^{15}$~cm$^{-3}$, Table~S1) in MgSc$_2$Se$_4$ may not significantly affect the XRD pattern{\cite{CanepaBoGopalakrishnanEtAl2016}} with respect to an ideal spinel structure and might be hard to detect using bulk characterization experiments. Also, the $c[e]^{eq}\sim$~7.96$\times$$10^{15}$ in MgSc$_2$Se$_4$ (Table~{\ref{tb:defects}}), corresponding to $\sim$~0.0001 free electrons per lattice site, is remarkably high compared to the intrinsic carrier concentration of typical semi-conductors (e.g., $\sim$~10$^{10}$~cm$^{-3}$ in Si), but significantly below metallic levels ($\sim$~1e$^{-}$ per lattice site). Although the In$^{\bullet}_{\rm Mg}$ and Mg$^{'}_{\rm In}$ defects charge-compensate in MgIn$_2$S$_4$, the $E_{Fermi}^{eq}$ under S-poor conditions is only $\sim$~0.2~eV below the CBM, indicating significant $n$-type conductivity. Indeed, a previous measurement of the Hall effect  in MgIn$_2$S$_4${\cite{Wakaki1980}} reported a moderate resistivity of $\sim$~8.2$\times$10$^{3}$ $\Omega$/cm and a free-electron concentration of $\sim$~6.4$\times$10$^{15}$ cm$^{-3}$, in  reasonable agreement with our $c[e]^{eq}$ estimate of $\sim$~4.1$\times$$10^{14}$~cm$^{-3}$  in the $\gamma$ facet. 

\subsection{Impact of cooling rates}

The variation of $c[e/h]^{frozen}$ and  $E_{Fermi}^{frozen}$ as a function of quench temperature ---the temperature at which the defect concentrations are frozen--- is plotted in Figures~{\ref{fig:summary}} and S7, respectively. Solid and dashed lines in Figure~{\ref{fig:summary}} correspond to anion-rich and anion-poor conditions, while the blue, red, and green colors indicate MgSc$_2$Se$_4$, MgIn$_2$S$_4$, and MgSc$_2$S$_4$.  The quench temperature, which is determined by the cooling rate significantly impacts the hole/electron conductivity. For example, all three spinels are expected to show spontaneous $h^+$ conductivity at 300~K in the anion-rich domain ($\alpha$ facet) when defect concentrations are quenched from 1300~K, contrary to the equilibrium scenario which would give negligible $p$\st{$/n$}-type conduction, as indicated by Figure~{\ref{fig:summary}} and Table~{\ref{tb:defects}}. Furthermore, quenched defect conditions in the anion-poor domain ($\gamma$ facet) dramatically increase the $n$-type conductivity in all spinels, resulting in $c[e]^{frozen}$ that are $\approx$~3 to 4 orders of magnitude higher than $c[e]^{eq}$ (Table~{\ref{tb:defects}}, Figure~{\ref{fig:summary}}). As a result, the synthesis of the chalcogenide spinels discussed in this work not only requires anion-rich conditions but also slow cooling post-synthesis (i.e., low quench temperatures, $\sim~400-500$~K, see Figure~S7) to minimize the electronic conductivity.

\begin{figure}[!ht]
\includegraphics[width=\columnwidth]{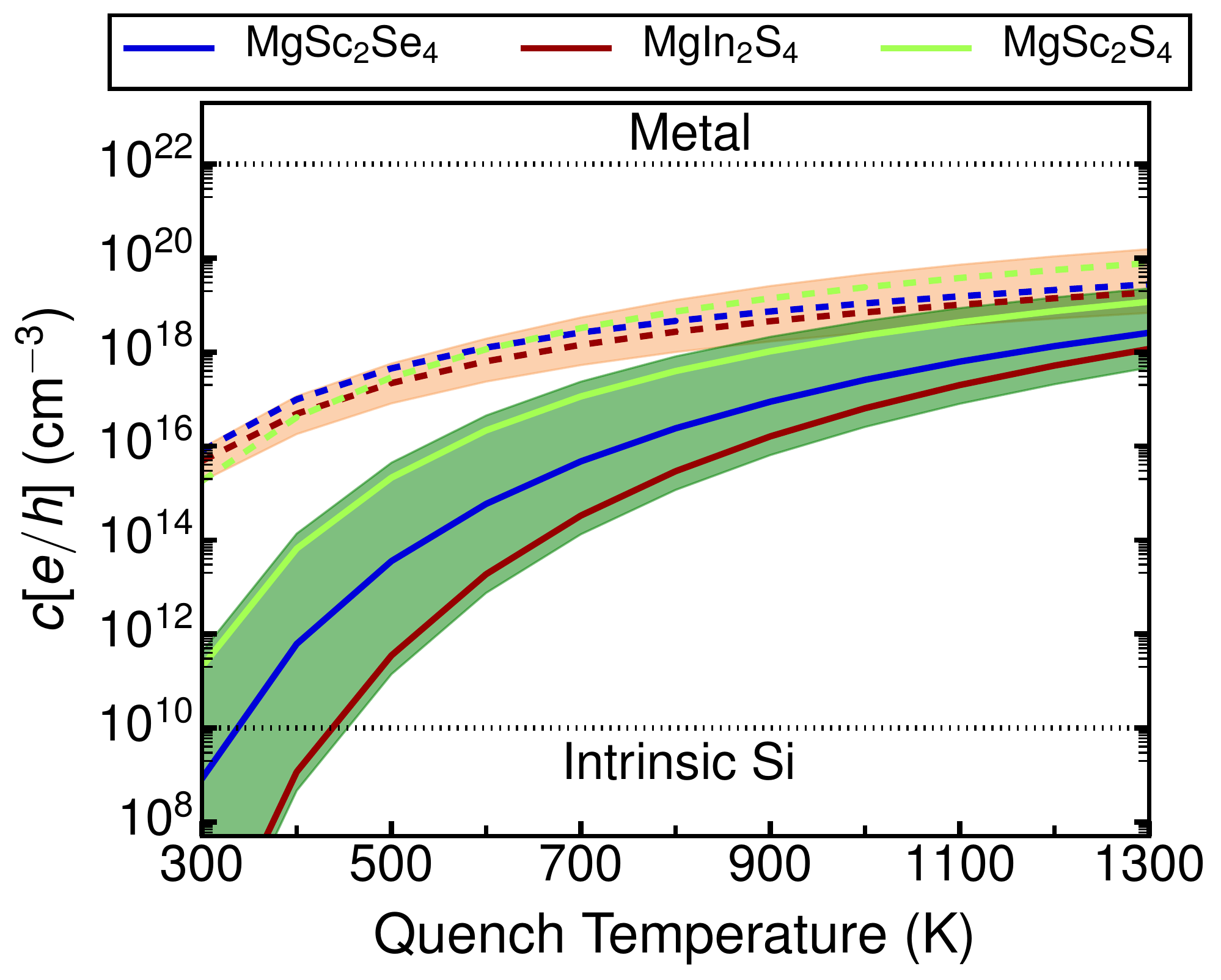}
\caption{
\label{fig:summary}  Free electron or hole concentrations $c[e/h]$ at 300~K as a function of temperature at which defect concentrations are quenched from. Solid and dashed lines indicate anion-rich (green-shaded) and anion-poor (orange-shaded) regions, respectively. The blue, red, and green line colors correspond to MgSc$_2$Se$_4$, MgIn$_2$S$_4$, and MgSc$_2$S$_4$. The dotted black lines indicate the typical free carrier concentration in intrinsic Si (10$^{10}$~cm$^{-3}$) and in metals (10$^{22}$~cm$^{-3}$). The $y$-axis values at 300~K are the $c[e/h]^{eq}$ for each spinel and values at 1300~K should indicate $c[e/h]^{frozen}$ corresponding to quenched defect concentrations from typical synthesis temperatures.{\cite{CanepaBoGopalakrishnanEtAl2016}} For the case of anion-rich MgIn$_2$S$_4$, the free carrier concentration is taken as the maximum of free electron and hole concentrations at each quench temperature.}
\end{figure}

\subsection{Inversion in MgIn$_2$S$_4$}
In comparison to the Sc-compounds, the defect energies in MgIn$_2$S$_4$ (Figure~\ref{fig:MgInSdefects}a and b) dictate that under equilibrium, the $E_{Fermi}^{eq}$ should be largely set by charge-compensating In$^{\bullet}_{\rm Mg}$, Mg$^{'}_{\rm In}$, and Vac$^{''}_{\rm Mg}$, corresponding to a lower hole/electron conductivity in either S-rich or S-poor condition.  Notably, the combination of In$^{\bullet}_{\rm Mg}$ and Mg$^{'}_{\rm In}$ anti-sites leads to inversion in the spinel (i.e., Mg and In exchange their respective sites), resulting in a [Mg$_{1-i}$In$_i$][Mg$_i$In$_{2-i}$]S$_4$ stoichiometry, where $i$ is the degree of inversion.  Our calculations indicate that MgIn$_2$S$_4$ will display significant spinel inversion under both S-rich and S-poor equilibrium conditions, with expected concentrations of 4.9$\times$$10^{17}$  for both In$^{\bullet}_{\rm Mg}$ and Mg$^{'}_{\rm In}$ (Table~S1), which qualitatively agrees with experimental reports.\cite{Wakaki1980,Burdett1981,Burdett1982,Sickafus2004,Zhang2010,Lucero2011} Spinel inversion can impact Mg-mobility and in turn the overall ionic conductivity since inverted structures will possess multiple local Mg--In configurations. 

Interestingly, the Sc-containing spinels are not expected to invert as much as the MgIn$_2$S$_4$. For example, MgSc$_2$S$_4$ exhibits fewer anti-sites (Sc$^{\bullet}_{\rm Mg}$, Mg$^{'}_{\rm Sc}$ $\sim~10^{11}$~cm$^{-3}$) than MgIn$_2$S$_4$ under S-rich equilibrium conditions (Table~S1). The tendency of MgIn$_2$S$_4$ to invert readily may be due to the $sp^{3}$ hybridization in the tetrahedra  that is better accommodated by In  than by Sc.

\subsection{Chemical driving forces for anti-site defect formation}
In the previous sections and Table~{\ref{tb:defects}}, we have demonstrated that the dominant defects in the chalcogenide spinels are anti-sites, signifying that $p$-type Mg$^{'}_{\rm In}$ and $n$-type In$^{\bullet}_{\rm Mg}$ are the primary intrinsic defects in MgIn$_2$S$_4$, while Sc$^{\bullet}_{\rm Mg}$ or Mg$^{'}_{\rm Sc}$ are the main defects in MgSc$_2$S$_4$(Se$_4$). These findings are similar to previous computational{\cite{Stevanovic2010,Paudel2011,Stevanovic2011,Das2016}} and experimental{\cite{Sickafus2004}} studies on ternary oxide spinels, with anti-sites dominating over other intrinsic defects, i.e. vacancies and interstitials. Given that the $p$-type ($n$-type) anti-site can compensate the excess electron (hole) ionized from the oppositely charged $n$-type ($p$-type) anti-site, the resulting Fermi level and the concentration of free electrons (or holes) at equilibrium depends on the difference in concentration between the $p$- and $n$-type anti-sites. For example, our data shows that under S-rich conditions an equal concentration of $p$- and $n$-type defects (as indicated by $E_{Fermi}^{eq}$ and $c[e/h]^{eq}$ in MgIn$_2$S$_4$, Table~{\ref{tb:defects}} and Table~S1), will pin the Fermi level within the band gap corresponding to a low concentration of free carriers.

So far, our calculations suggest that anti-site concentrations (and the corresponding difference between the concentration of $p$- and $n$-type anti-sites) can be markedly different for the spinels considered in this work. Particularly, the results presented in Table~{\ref{tb:defects}} demonstrate that the concentration of anti-sites in MgIn$_2$S$_4$ is always greater by several orders of magnitude (across all chemical conditions, see Table~S1) than in MgSc$_2$S$_4$(Se$_4$). Additionally, our calculations indicate that the difference between the $p$- and $n$-type anti-site concentrations in MgIn$_2$S$_4$ is consistently lower than the Sc-spinels (Table~S1), with profound effects on the type and magnitude of the electrical conductivity in the corresponding systems. Here, we rationalize the chemical factors driving such differences.

In general, the occurrence of anti-sites depends on a combination of several factors, such as, $i$) \emph{steric effects} (i.e. the strain due to differences in ionic radii of the cations forming anti-sites), $ii$) the \emph{band-gap} of each material (ease of ionizing the excess electron/hole), and $iii$) the electronic nature or \emph{bond character} of specific bonds (covalent or ionic).

\renewcommand{\labelenumi}{\roman{enumi}}
\begin{enumerate}

\item {\emph{Steric effects}: anti-sites are facilitated if the cations substituting for each other possess similar ionic size.  For example, Mg$^{2+}$ has an ionic radius of $\sim$~0.57~{\AA{}} and $\sim$~0.72~{\AA{}} in tetrahedral and octahedral coordination, respectively, which compares well with the ionic radius of In$^{3+}$ $\sim$~0.62~{\AA{}} in tetrahedral and $\sim$~0.80~{\AA{}} in octahedral coordination, respectively,{\cite{Shannon1976}} implying the facile formation of both Mg$^{'}_{\rm In}$ and In$^{\bullet}_{\rm Mg}$ anti-sites. While Sc$^{3+}$ has an ionic radius of $\sim$~0.75~{\AA{}} in octahedral sites,{\cite{Shannon1976}} it has never been observed in tetrahedral coordination to our knowledge.
}

\item {\emph{Band-gap}: large band gaps in materials limit the possibility of ionization of the excess charge in defects, penalizing the injection of a free hole (electron) into the valence (conduction) band. Thus, the ``large" band gap in MgIn$_2$S$_4$ (Figure~S3 in the SI) indicates a high energy penalty to ionize the excess charge, forcing Mg$^{'}_{\rm In}$ to charge compensate In$^{\bullet}_{\rm Mg}$ (and vice-versa) and leading to a lower difference in concentration between Mg$^{'}_{\rm In}$ and In$^{\bullet}_{\rm Mg}$ across all conditions (Table~S1). In contrast, the band gaps in Sc-spinels are quantitatively lower than the In-spinel, indicating that the energy penalty for either Sc$^{\bullet}_{\rm Mg}$ or Mg$^{'}_{\rm Sc}$ to ionize the excess charge is significantly smaller, suggesting that the Mg$^{'}_{\rm Sc}$ may not be required to charge compensate the Sc$^{\bullet}_{\rm Mg}$. Since inversion in the spinel structure correlates with the formation of comparable quantities of both $p$- and $n$-type anti-sites, MgIn$_2$S$_4$ is more susceptible in exhibiting spinel inversion than the Sc-compounds, in agreement with previous experimental studies.{\cite{Wakaki1980}} 
}

\item {\emph{Bond character}: covalent bonds with significant hybridization of the transition metal and the anion can tolerate anti-sites better than ionic bonds, due to greater electrostatic screening of the excess charge in the anti-sites. Since each octahedral ($16d$) site in the spinel structure shares edges with 6 other $16d$ sites (Figure~{\ref{fig:pdstructure}}),{\cite{Gautam2017a}} electrostatic screening will be important in stabilizing the $p$-type anti-sites (i.e., Mg$^{'}_{\rm In}$ and Mg$^{'}_{\rm Sc}$). From a qualitative analysis of the valence band edge in the density of states (Figure~S3 in the SI) in MgSc$_2$S$_4$(Se$_4$), we speculate that both Mg and Sc bond quite ionically with the anion (S/Se). In contrast, the In-S bonds show significant hybridization in MgIn$_2$S$_4$ compared to the Sc-S(Se) bonds in MgSc$_2$S$_4$(Se$_4$), stabilizing the $p$-type Mg$^{'}_{\rm In}$. These observations could explain the higher concentrations of Mg$^{'}_{\rm In}$ as opposed to Mg$^{'}_{\rm Sc}$, as indicated by our calculations across all conditions (Table~S1).
}

\end{enumerate}

\noindent From this analysis, two criteria to design a ternary spinel ionic conductor with minimal electronic conductivity emerge: $i$) materials with large band-gap (curbing the ionization of free carriers) and $ii$) materials where both $p$- and $n$-type anti-sites are equally likely to form (leading to spinel inversion and lower free carriers) are preferable.

\subsection{Extrinsic doping}
Aliovalent doping of ionic conductors can be used as a strategy to both enhance the ionic conductivities while suppressing intrinsic electronic (hole) conductivities. We explored the defect chemistry of extrinsic dopants in the $\gamma$ facet of MgSc$_2$Se$_4$ (Figure~{\ref{fig:extrinsics}), comprising tetravalent and pentavalent non-redox cation substitution on Sc as well as halogen doping on Se. Given that the intrinsic $E_{Fermi}^{eq}$ is $\sim$~1.08~eV in anion-poor MgSc$_2$Se$_4$, doping of most cations is not energetically favored (Figure~{\ref{fig:extrinsics}}a), with the exception of Ce, Ti, and Zr. Halogen doping on Se appears feasible ($E_f < 0.4$~eV, Figure~\ref{fig:extrinsics}b), although it may further increase  the $n$-type behavior. In contrast, $n$-type Ce and Ti have their respective donor transition levels deeper in the band-gap (Figure~{\ref{fig:extrinsics}}a) and may reduce the free electron concentration in MgSc$_2$Se$_4$. However, it remains to be experimentally  confirmed whether Ti and Ce can be efficiently doped on the Sc site.

\section{Conclusion}
Using first-principles calculations, we have analyzed the role of defect chemistry in influencing the electrical conductivities of three chalcogenide spinels, MgSc$_2$Se$_4$, MgSc$_2$S$_4$ and MgIn$_2$S$_4$, which are potential Mg-ion conductors. We find that intrinsic point defects, such as Mg-metal anti-sites (\{Sc/In\}$^{\bullet}_{\rm Mg}$, Mg$^{'}_{\rm \{Sc/In\}}$) and Mg-vacancies (Vac$^{''}_{\rm Mg}$), dramatically affect the free carrier concentrations of the spinels under consideration. Additionally, controlling the anion-content during synthesis is an important factor in determining the defect energetics and the resultant electrical conductivity, with all three spinels exhibiting high $n$-type conductivity in anion-poor conditions and marginal $p$-type behavior in anion-rich conditions. Also, fast cooling leads to large concentrations of intrinsic defects being quenched within the structure, which can  increase both the free hole (anion-rich) and free electron (anion-poor) concentrations in MgSc$_2$Se$_4$, MgSc$_2$S$_4$ and MgIn$_2$S$_4$. Hence, the lowest electronic conductivity is to be expected for samples synthesized under anion-excess, and slowly cooled to room temperature. Among the three structures considered, MgIn$_2$S$_4$ exhibits the lowest free carrier concentration across various conditions, largely due to inversion within the spinel. Finally, the introduction of aliovalent dopants, such as Ce and Ti on Sc, may mitigate the electronic conductivity observed in MgSc$_2$Se$_4$. Our work indicates the importance of defects in the field of solid electrolytes, and the framework used here can be applied to other systems as well, which will eventually aid both in the calibration of existing candidates and accelerated materials discovery.


\begin{acknowledgement}
This work was fully supported as part of the Joint Center for Energy
Storage Research (JCESR), an Energy Innovation Hub funded by the U.S.
Department of Energy, Office of Science, and Basic Energy Sciences. This
study was supported by Subcontract 3F-31144.  The authors thank the National Energy Research Scientific Computing Center (NERSC) for providing computing resources.  The authors thank Mark D.\ Asta at UC Berkeley for constructive comments, while PC acknowledges Jacques--Arnaud D\^{a}wson for useful discussion. 
\end{acknowledgement}
%%%%%%%%%%%%%%%%%%%%%%%%%%%%%%%%%%%%%%%%%%%%%%%%%%%%%%%%%%%%%%%%%%%%%%%%

%%%%%%%%%%%%%%%%%%%%%%%%%%%%%%%%%%%%%%%%%%%%%%%%%%%%%%%%%%%%%%%%%%%%%%%%
% References and Notes
\bibliography{biblio}
%%%%%%%%%%%%%%%%%%%%%%%%%%%%%%%%%%%%%%%%%%%%%%%%%%%%%%%%%%%%%%%%%%%%%%%%

%%%%%%%%%%%%%%%%%%%%%%%%%%%%%%%%%%%%%%%%%%%%%%%%%%%%%%%%%%%%%%%%%%%%%%%%
\end{document}